\documentclass[prl,twocolumn,showpacs,preprintnumbers,superscriptaddress,amsmath,amssymb]{revtex4}
\usepackage{graphicx}
\usepackage{dcolumn}
\usepackage{bm}

\begin{document}

\title{Coherent intense resonant laser pulses lead to interference in the time domain observable in the spectrum of the emitted particles}

\author{\firstname{Philipp~V.} \surname{Demekhin}}
\altaffiliation{philipp.demekhin@pci.uni-heidelberg.de}
\affiliation{Theoretische Chemie, Physikalisch-Chemisches Institut, Universit\"{a}t  Heidelberg, Im Neuenheimer Feld 229,
D-69120 Heidelberg, Germany}
\affiliation{Rostov State Transport University, Narodnogo Opolcheniya square 2,  Rostov-on-Don, 344038, Russia}

\author{\firstname{Lorenz~S.} \surname{Cederbaum}}
\affiliation{Theoretische Chemie, Physikalisch-Chemisches Institut, Universit\"{a}t Heidelberg,
Im Neuenheimer Feld 229, D-69120 Heidelberg, Germany}

\date{\today}

\begin{abstract}
The dynamics of atomic levels resonantly coupled by a coherent and intense short high-frequency laser pulse is discussed and it is advocated that this dynamics is sensitively probed by measuring the spectra of the particles emitted. It is demonstrated that the time-envelope of this laser pulse gives rise to two  waves emitted with a time delay with respect to each other at the rising and falling sides of the pulse, which interfere in the time domain. By computing numerically and analyzing explicitly analytically a show-case example of sequential two-photon ionization of an atom by resonant laser pulses, we argue that this dynamic interference should be a general phenomenon in the spectroscopy of strong laser fields. The emitted particles do not have to be photoelectrons. Our results allow also to interpret the already studied resonant Auger effect of an atom by intense free electron laser  pulses, and also to envisage experiments in which photons are emitted.
\end{abstract}

\pacs{33.20.Xx,  41.60.Cr, 82.50.Kx}

\maketitle

The interaction of an atom with intense laser fields has been widely studied. If the field is essentially monochromatic, the physics is well described by a time-independent Hamiltonian in the basis of 'dressed' electronic states or Floquet states (see, e.g., Refs.~\cite{B1,B2,B3,B4}). The inclusion of relaxation mechanisms, such as autoionization or subsequent ionization, gives a 'dressed' state a finite width, and it becomes unstable \cite{NIMROD}. The concept of dressed states is applied in practically every branch of spectroscopy of optical lasers operating in the nano and picosecond regimes. If the laser pulses are shorter, a Floquet basis is still useful, but one has to take the time-dependence of the pulse explicitly into account. Many new phenomena arise due to the impact of this time-dependence \cite{ZEWAIL,KRAUSZ,THbp1,THbp2,THbp3}.

One class of such phenomena extends the well-known stationary Rabi-doublets existing in strong fields owing to ac-Stark splitting or Autler-Townes effect \cite{AT}. Because of the short optical pulse, the value of this splitting varies, resulting in the appearance of a multiple-peak interference pattern in the computed autoionization \cite{AUI1} and resonant multiphoton ionization  \cite{MPI1,MPI2} electron spectra. This new pattern is attributed to the temporal coherence of a pulse strong enough to induce Rabi oscillations between resonantly coupled states \cite{AUI1,MPI1,MPI2}. However, a physically simple explanation of the phenomenon is still missing \cite{Rongqing}.

Although these theoretical predictions are of relevance and were made a long time ago, they have not been verified experimentally so far. To our opinion, this is due to the optical regime. First, in this regime there are rarely well separated resonances and there is often a dense spectrum of close-by Rydberg and doubly-excited states which also participate in the dynamics. Second, these states induce additional ac-Stark shifts which vary in time \cite{Sussman11}. Third, one is often in the vicinity of ionization thresholds and ionization is particularly efficient there. All of these additional states and effects strongly smear out the pronounced interference pattern which would be obtained if only two or three states were resonantly coupled by the pulse.

The situation becomes particularly promising by the advent of the new generation of light sources, like attosecond lasers  \cite{KRAUSZ}, high-order harmonic generation sources \cite{highharm1,highharm2}, and free electron lasers \cite{FLASH,FERMI} to produce ultrashort and intense coherent laser pulses of high frequencies. The above mentioned shortcomings which impede experimental verifications by optical pulses, are absent at higher frequencies and one can study the dynamics of a few well separated electronic states (e.g., core-excited states) resonantly coupled by a short coherent pulse. Unless the intensity is very high, the resonant dynamics will not be affected by ac-Stark shifts arising from   nonessential states, and the impact of direct ionizations is not substantial since the photoionization probability usually decreases with the photon energy. We thus concentrate in this work on the high-frequency regime and discuss a fundamental consequence of the nature of intense coherent laser pulses on spectroscopic observables. Due to the high carrier frequencies, much of the physics follows the evolution provided by the pulse envelope nearly adiabatically up to rather short pulse durations. This makes the underlying physics particularly transparent.

Let us consider two bound electronic states of an atom coupled resonantly by a strong laser pulse (the carrier frequency can be different from the field-free resonant frequency to compensate for the emerging energy detuning by the AC Stark effect). The two initially degenerate `dressed'  states repel each other by the field-induced coupling and split in energy. If the pulse envelope supports many optical cycles of high frequency, the field-induced coupling between the two electronic states adiabatically follows the pulse envelope \cite{Sussman11}. Consequently, the energy splitting will adiabatically increase when the pulse arrives and then decrease when the pulse expires. If the atom emits particles during its exposure to the pulse (photoelectrons, Auger electrons, photons), it will become evident below that the particles emitted when the pulse rises have the same kinetic energy as those emitted  when the pulse decreases. The respective two waves emitted with a time delay with respect to each other will interfere and their spectrum will exhibit a pronounced interference pattern. We would like to call this kind of interference, dynamic interference.

Although we concentrate here on high-frequency short pulses coupling two bound states, we mention that bound-continuum coupling by such pulses also leads to dynamic interference in the ionization spectra of atoms \cite{DynIntLETT} and model anions \cite{NOTOURS}. Furthermore, oscillations in the total multiphoton ionization yield as function of laser intensity have been observed for atoms exposed to optical lasers \cite{EXPT,jones} and interpreted as arising from interferences of electrons emitted at different times \cite{jones}. We shall demonstrate here that dynamic interference is a general consequence of the finite nature of intense high-frequency laser pulses, and leads to pronounced patterns observable in the spectrum of the emitted particles. We first concentrate on a show-case example of sequential two-photon ionization of an atom by strong pulses. The example is of much interest by itself, since the coupled two-level system is probed here by a second photon of the same pump pulse. Our results pave the way for experiments on dynamic interference by available laser pulse sources. We also briefly discuss the effect of dynamic interference in other branches of laser spectroscopy.

Below, we consider an atom initially in its ground electronic state $\vert I\rangle$ of energy of $E_I=0$ chosen as the origin of the energy scale, which is resonantly excited into the intermediate state $\vert R\rangle$ of energy $E_R$  by absorption of a single photon and subsequently ionized by a second photon into a final electron continuum state $\vert F \varepsilon\rangle$ of energy $IP + \varepsilon$. Here, $IP=E_F-E_I$ is the ionization potential and $\varepsilon$  is the kinetic energy of the emitted photoelectron. Employing a linearly polarized coherent laser pulse
$\mathcal{E}(t)=\mathcal{E}_0 \,g(t) \cos\omega t$ with  pulse envelope $g(t)$, the total wave function  as a function of time reads \cite{Demekhin11SFatom,DynIntLETT,Chiang10,MolRaSfPRL,DemekhinCO,DemekhinICD}
\begin{equation}
\label{eq:anzatzR}
\Psi(t)= a_I(t)\vert I \rangle+ {a}_R(t)e^{-i\omega t} \vert R \rangle+\int  {a}_\varepsilon (t ) e^{-2i\omega t}\vert F \varepsilon \rangle d\varepsilon.
\end{equation}
where $a_I(t)$, ${a}_R(t)$, and ${a}_\varepsilon (t)$ are the time-dependent amplitudes for the population of the  $\vert I \rangle$, $\vert R \rangle$, and  $\vert F \varepsilon \rangle$ levels, respectively. The stationary states $\vert R \rangle$  and  $\vert F \varepsilon \rangle$ have been `dressed' by multiplying with the phase factors  $e^{i\omega t}$ and $e^{2i\omega t}$ \cite{Demekhin11SFatom}, to simplify the equations of motion.

Inserting $\Psi(t)$ into the time-dependent Schr\"{o}dinger equation for the total Hamiltonian, and implying also the rotating wave \cite{THbp2,THbp3} and local \cite{Demekhin11SFatom,Cederbaum81,Domcke91} approximations, we obtain the following set of equations for the amplitudes (atomic units are used throughout)
\begin{subequations}
\label{eq:CDE}
\begin{equation}
\label{eq:CDE_I}
i\dot{a}_I(t)=  \frac{ D^\dag  \mathcal{E}_0}{2}  g(t)  {a}_R(t),
\end{equation}
\vspace{-0.75cm}
\begin{equation}
\label{eq:CDE_F}
i\dot{a}_R(t)=   \frac{ D  \mathcal{E}_0}{2} g(t) \, {a}_I(t) + (E_R -\begin{array}{c}   \frac{i}{2}\Gamma g^2(t) \end{array} -\omega)  a_R(t),
\end{equation}
\vspace{-0.75cm}
\begin{equation}
\label{eq:CDE_G}
i\dot{ {a}}_\varepsilon (t)=   \frac{d \mathcal{E}_0}{2}   a_R(t) + \left(IP+ \varepsilon-2\omega\right) {a}_\varepsilon (t).
\end{equation}
\end{subequations}
Here, $D=\langle R\vert \hat{z}\vert I\rangle$ and $d=\langle F\varepsilon \vert \hat{z}\vert R\rangle$ are the dipole transition matrix elements for the excitation of the intermediate state and for its subsequent ionization, respectively. The term $-\frac{i}{2}\Gamma g^2(t)$ in Eq.~(\ref{eq:CDE_F}) is the time-dependent ionization rate of the intermediate state responsible for the leakage of its population by the ionization into all final continuum states $\vert F\varepsilon\rangle$, turning this state into a resonance. Explicitly, $\Gamma = 2\pi \vert  d   \mathcal{E}_0  /2 \vert^2 $  \cite{Demekhin11SFatom,Sun}.

\begin{figure}
\includegraphics[scale=0.32]{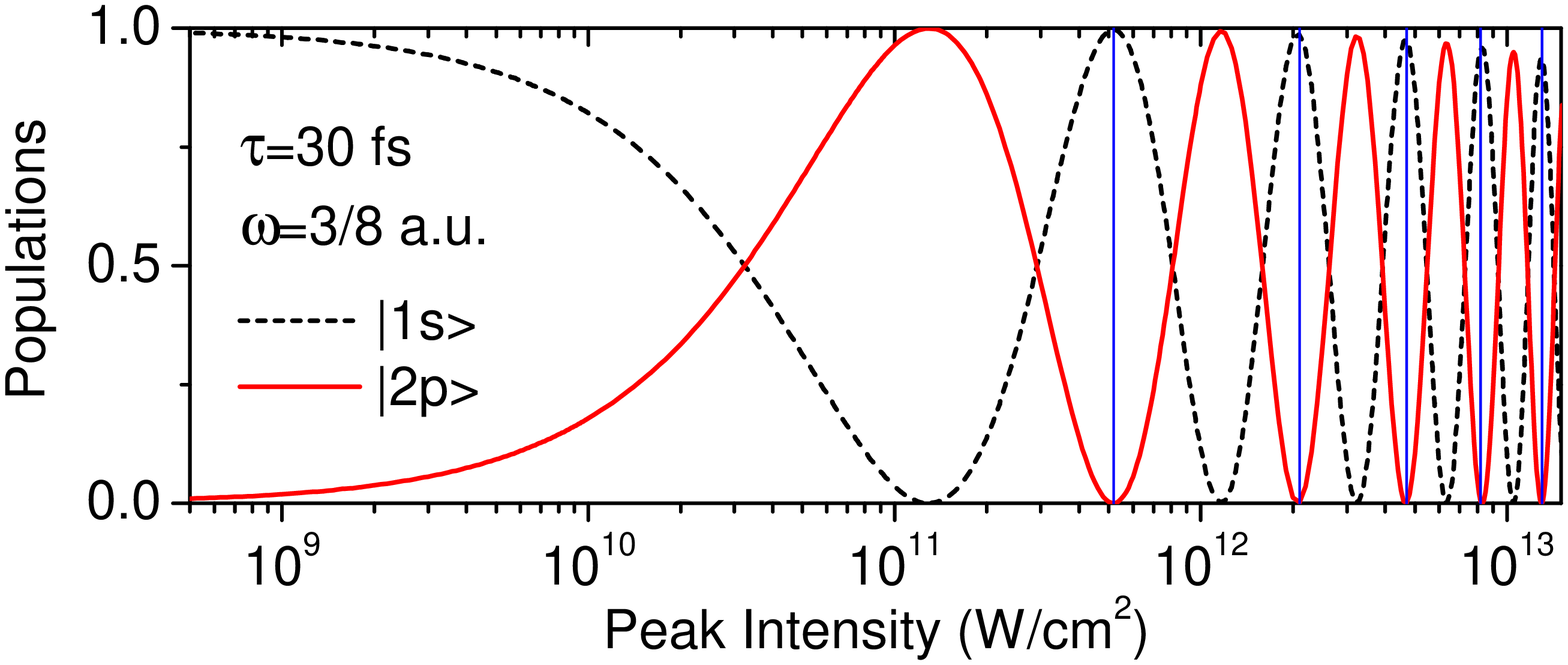}
\caption{(Color online) Sequential two-photon ionization of H by a Gaussian-shaped pulse of 30 fs duration and resonant carrier frequency of $\omega=3/8$~a.u.=10.20~eV, which fits to the energy of the H(1s)--H(2p) excitation. Shown are the populations of the ground state H(1s) and of the resonant state H(2p) as functions of the peak intensity after the laser pulse has expired. The vertical lines indicate the peak intensities at which the  spectra depicted in Fig.~\ref{fig_spe} are computed. }\label{fig_pop}
\end{figure}

To exemplify the present theory, we study the sequential two-photon ionization of the hydrogen atom. In the process, H(1s) is resonantly excited to H(2p) state, which is then ionized. The photon energy was set to fit the excitation energy $\omega=E_R=3/8~\mathrm{a.u.}=10.20~\mathrm{eV}$. The computed dipole transition matrix elements for the excitation and ionization are $D=0.744$~a.u.  and $d=0.377$~a.u., respectively. The system of Eqs.~(\ref{eq:CDE}) was solved numerically employing a Gaussian pulse $g(t)=e^{-t^2/\tau^2}$ of $\tau=30$~fs duration.  Fig.~\ref{fig_pop} shows the populations of the ground state H(1s) and of the resonant state H(2p) after the laser pulse has expired as function of the peak intensity $I_0=\mathcal{E}^2_0/8\pi\alpha$. The populations exhibit pronounced Rabi oscillations. To be noticed is that at the highest intensity considered in Fig.~\ref{fig_pop}, the total photoelectron yield reaches just 7\% indicating that the ionization by the second photon is far from saturation.

\begin{figure}
\includegraphics[scale=0.42]{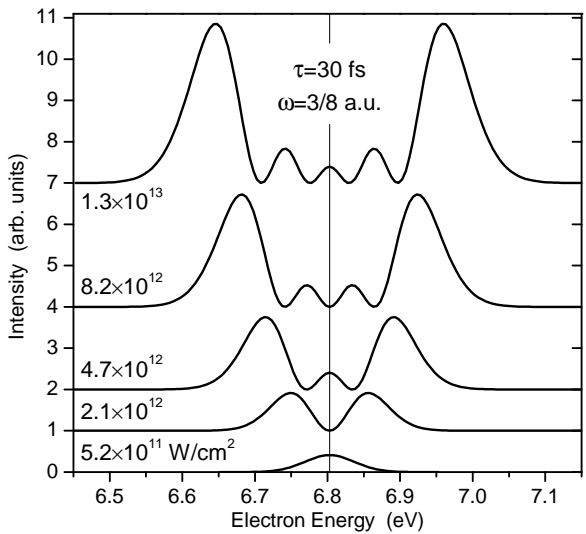}
\caption{Sequential two-photon ionization of H by a Gaussian-shaped pulse of 30 fs duration and resonant carrier frequency of $\omega=E_R=3/8~\mathrm{a.u.}=10.20~\mathrm{eV}$, which fits to the energy of the H(1s)--H(2p) excitation. Shown are the photoelectron spectra computed via Eqs.~(\ref{eq:CDE}) for different peak intensities indicated in the figure near each curve. The central electron energy $\varepsilon_0=2 \omega - IP=0.25~\mathrm{a.u.}=6.80~\mathrm{eV}$ at which the photoelectron spectrum has its maximum in the weak-field case is indicated by a vertical line. }\label{fig_spe}
\end{figure}

We now turn to the photoelectron spectra. For the calculations we have chosen the peak intensities at the maxima of the ground state population indicated in  Fig.~\ref{fig_pop}  by vertical lines. At these intensities the atom manages to complete an integer number of Rabi cycles during the pulse duration. The spectra computed via  Eqs.~(\ref{eq:CDE}) are shown in Fig.~\ref{fig_spe}.  The spectrum computed for the lowest considered intensity of $5.2\times 10^{11}$~W/cm$^2$ is rather close to that expected in the weak-field case, i.e., a Gaussian curve
centered around $\varepsilon_0=2 \omega - IP=0.25~\mathrm{a.u.}=6.80~\mathrm{eV}$. As the field intensity increases and the atom manages to complete two Rabi cycles while the pulse is on (second spectrum from the bottom), the spectral distribution bifurcates, and is now minimal at $\varepsilon_0$. At the intensity $4.7\times 10^{12}$~W/cm$^2$  when  the atom has completed three Rabi cycles,  the spectrum  bifurcates again and possesses now three maxima (third spectrum from the bottom). As the pulse intensity grows further, the spectrum continues to bifurcate again and again, exhibiting thereby distinct multiple-peak structures.   Below we identify dynamic interference as the physical origin of these patterns.

To start the discussion, we notice that the resonantly ($\omega=E_R-E_I$) coupled  dynamics of the $\vert I\rangle$ and  $\vert R\rangle$ states in Eqs.~(\ref{eq:CDE_I}) and (\ref{eq:CDE_F}) is  governed by the $2\times 2$ Hamiltonian
\begin{equation}
\label{eq:ham}
\mathbf{H}(t)=  \left[ \begin{array}{cc} 0&     \Delta^\dag\,g(t)\\ \Delta\,g(t)& -   \frac{ i }{2}\Gamma g^2(t)    \end{array}   \right],
\end{equation}
where $\Delta=\frac{ D  \mathcal{E}_0}{2}$. The ionization of the intermediate state by a second photon from the same pulse is described in Eq.~(\ref{eq:ham}) by the $-\frac{ i }{2}\Gamma g^2(t)$ term,  and actually probes  this Hamiltonian. We may now follow the time evolution of the eigenvalues and eigenvectors of this Hamiltonian.

When the pulse is  on, the solution of Eq.~(\ref{eq:ham}) yields two decoupled resonances, which are superpositions of the initial $\vert I\rangle$  and intermediate $\vert R\rangle$ states:
\begin{equation}
\label{eq:supep}
E_{\pm}(t)\simeq   \pm \Delta\,g(t)-  \frac{ i }{4}\Gamma g^2(t), ~~\vert \pm\rangle \simeq \dfrac{\vert I\rangle}{\sqrt{2}} \pm\dfrac{\vert R\rangle}{\sqrt{2}}.
\end{equation}
This result is well justified if the pulse is not too strong and the ionization is far from saturation, i.e., when $\Delta\,g(t) \gg \frac{1 }{2} \Gamma g^2(t) $. These solutions describe two decoupled time-independent resonances with time-dependent energies   $\pm \Delta\, g(t)$ induced by the field. Importantly, their energies move apart as the pulse arrives, and then move towards  each other  as the pulse expires. Both resonances are subject to the same leakage  $-\frac{i}{4}  \Gamma \,g^2(t)$, populating thereby the continuum states $\vert F \varepsilon \rangle$ via the ionization by a second photon.

The decoupled resonances scenario enables one to uncover the origin of oscillations in the spectra  in Fig.~\ref{fig_spe}. Using Eqs.~(\ref{eq:supep}) we can rewrite the original Eqs.~(\ref{eq:CDE}) in terms of the decoupled resonances $\vert +\rangle$ and $\vert -\rangle$  and obtain the  equations for the amplitudes $a_+(t)$ and $a_-(t)$ of these resonances which can be solved analytically.  Employing  the initial conditions $a_{\pm}(-\infty)={1}/{\sqrt{2}} $ we find
\begin{equation}
\label{eq:GSsol}
{a}_{\pm}(t)= \begin{array}{c}  \frac{1}{\sqrt{2}}  \end{array}e^{ \left[\mp i\Delta\,F(t)-  \Gamma/4  \, J(t)\right] },
\end{equation}
where   $F(t)=\int_{-\infty}^{t}g(t^\prime) dt^\prime$ and $J(t)=\int_{-\infty}^{t}g^2(t^\prime) dt^\prime$ are time-integrals over the pulse envelope and its square.

The population amplitudes  $a_\varepsilon(t)$ in Eq.~(\ref{eq:CDE_G}) can be expressed as an integral of $a_{R} (t)$ \cite{Demekhin11SFatom} and, after employing Eqs.~(\ref{eq:supep}), as an integral of  $a_{+} (t)-a_{-} (t)$. Using now the explicit expressions (\ref{eq:GSsol}) makes the computation of ${a}_\varepsilon (t)$ and of the spectrum $\sigma(\varepsilon)=\vert {a}_\varepsilon (\infty) \vert^2 $ rather straightforward
\begin{multline}
\label{eq:statphase1}
\sigma(\varepsilon)=   \left|\frac{d \mathcal{E}_0}{4} \int_{-\infty}^\infty   g(t) \, e^{ -\Gamma/4\, J(t)} \right. \times \\ \left. \left\{ - e^{ i\left[\delta t+ \Delta\, F(t) \right]}+ e^{ i\left[\delta t- \Delta\,F(t) \right]}   \right\}  dt \right|^2,
\end{multline}
where we introduced the abbreviation $\delta= IP+\varepsilon -2\omega =\varepsilon - \varepsilon_0 $, which is the electron energy detuning from the center of the photoelectron spectrum $\varepsilon_0=2 \omega- IP$.

Interestingly, this expression  for the spectrum can further be evaluated analytically. To this end we notice that the  integrand in (\ref{eq:statphase1}) contains the sum of two rapidly oscillating factors which is multiplied by a smoothly varying function of time. The main contributions to the integral stem from the times at which two phases   $\Phi_{\pm}(t)=\delta_{\,}t\pm \Delta\,F(t)$ are stationary \cite{statphase}, i.e.  $\dot{\Phi}_{\pm}(t_s)=0$. The two resulting stationary time conditions, $\delta = \mp \Delta\,g(t_s) $, have a transparent physical meaning. They define the time $t_s(\varepsilon)$  at which an energy of a decoupled resonance, continuously shifted by the time-dependent coupling $\pm \Delta\, g(t)$, moves across the energy position $\delta=\varepsilon-\varepsilon_0$ of the continuum state under inspection. These times and the electron energy $\varepsilon$ are connected via the simple expression $\varepsilon=2 \omega- IP \mp\Delta\,g(t_s) $.  During the pulse  resonance $\vert -\rangle $  covers the lower kinetic electron energy side of the spectrum, $\varepsilon-\varepsilon_0 \in [-\Delta, 0]$, and resonance $\vert +\rangle $ the higher  energy side, $\varepsilon-\varepsilon_0 \in [ 0,+ \Delta]$.  For any pulse there are at least two stationary points for each value of $\varepsilon$: one, $t_1(\varepsilon)$, when the pulse is growing, and another, $t_2(\varepsilon)$, when it decreases. For a Gaussian pulse there are exactly two times,  $t_1(\varepsilon)=-t_2(\varepsilon) =\tau \sqrt{ \ln[\Delta/(\varepsilon-\varepsilon_0)]  }$.

\begin{figure}
\includegraphics[scale=0.42]{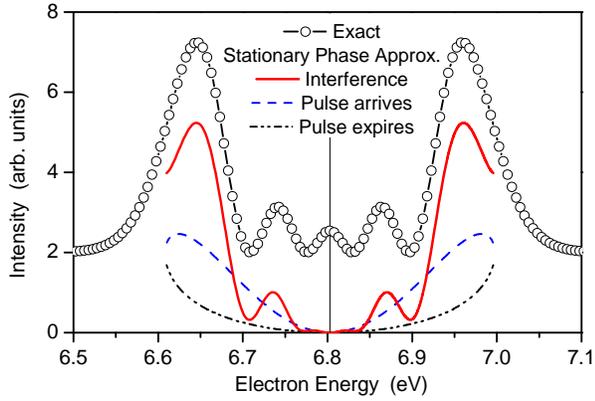}
\caption{(Color online) Sequential two-photon ionization of H by a Gaussian-shaped pulse of 30 fs duration, carrier frequency of $\omega=3/8$~a.u.=10.20~eV, and peak intensity of $1.3\times10^{13}$~W/cm$^2$. Shown are the photoelectron spectrum computed numerically (open circles; taken from Fig.~\ref{fig_spe}) and the spectrum obtained in the stationary phase approximation via the  explicit expression (\ref{eq:statphase}) (solid curve). The two individual contributions to the spectrum, describing in (\ref{eq:statphase}) the separate distributions of photoelectrons emitted at times when the pulse arrives and expires, are shown by broken curves.}\label{fig_app}
\end{figure}

By collecting in the integral (\ref{eq:statphase1}) the two stationary phase contributions at $t_s=\pm t_1(\varepsilon)$, we obtain the following explicit approximate expression for the spectrum
\begin{multline}
\label{eq:statphase}
\sigma(\varepsilon) \simeq \left| \frac{d \mathcal{E}_0}{4}  \sum_{t_s=\pm t_1(\varepsilon)} g(t_s) \,e^{ -\Gamma/4 J(t_s)} \right.\\
\left. \times \left\{ - e^{ i\left[ \Phi_+(t_s)\mp \frac{\pi}{4}  \right]}+ e^{ i\left[  \Phi_-(t_s)\pm\frac{\pi}{4}\right]}   \right\}       \right|^2.
\end{multline}
The additional phase factors $\frac{\pi}{4}$  result from higher terms in the expansion of the phase $\Phi_\pm(t)$ around the stationary points $\pm t_1(\varepsilon)$ computed for the Gaussian pulse \cite{DynIntLETT}. The photoelectron spectrum  Eq.~(\ref{eq:statphase}) is easily evaluated. The result is depicted in Fig.~\ref{fig_app} by a solid curve. It is illuminating to see that an explicit simple expression reproduces nicely the numerically determined spectrum (open circles). The individual contributions of the two times $t_s=\pm t_1(\varepsilon)$ to the spectrum  in  Eq.~(\ref{eq:statphase}) are rather smooth and do not show any interference effects (broken curves).

\begin{figure}
\includegraphics[scale=0.42]{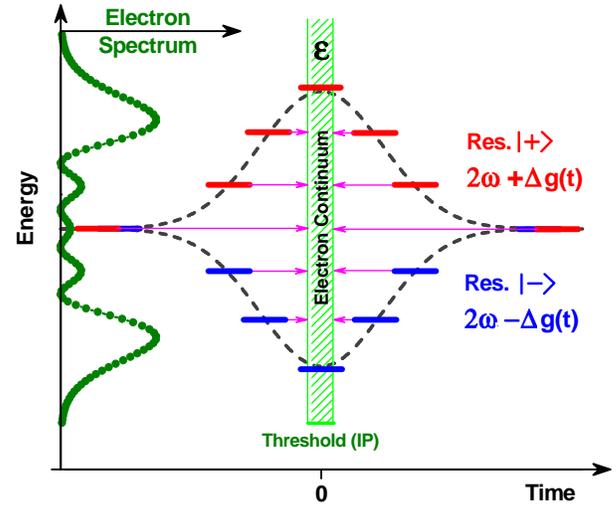}
\caption{(Color online) The intense laser pulse of resonant carrier frequency induces a time-dependent coupling $\Delta\,g(t)$ between the ground and intermediate states. The energies of the resulting two decoupled resonances follow adiabatically the pulse envelope $g(t)$ in two opposite directions (dashed curves). As a result, the photoelectron emitted by a second photon along the pulse envelope has at every moment $t$ predominantly a specific kinetic energy $\varepsilon$. These are the times at which the energies of the decoupled resonances move across the energy position $\delta=\varepsilon-\varepsilon_0$  of the continuum state. These passage times and the kinetic electron energy are simply connected via $\varepsilon=2 \omega- IP \mp\Delta\,g(t_s)  $. The pulse envelope first grows and then falls, and for a Gaussian pulse there are exactly two times at which the emitted electron wave has the same energy $\varepsilon$. These two waves emitted with a time delay with respect to each other interfere, giving rise to the strongly modulated distribution of the photoelectrons shown on the energy axis. Resonance $\vert -\rangle  $ is responsible for the low-energy\,part of the spectrum, and resonance $\vert +\rangle  $ for the high-energy part.}\label{fig_shm}
\end{figure}

Eq.~(\ref{eq:statphase})  uncovers the physical origin of the strong modulations in the electron spectrum. These are the results of the coherent superposition of two photoelectron waves emitted with the same kinetic energy at two different times. A schematic visualization of the dynamic interference is given in Fig.~\ref{fig_shm}. The dynamic interference spectacularly modifies the sequential two-photon ionization process and causes enormous qualitative changes in  the spectrum, which can be verified by available laser pulse sources. The predicted effect is not constrained to sequential two-photon ionization. We are convinced that dynamic interference is a very general and fundamental effect which is best manifested in the observable spectrum of the emitted particles by prominent multiple-peak patterns. Often, the dynamics of states coupled by intense laser pulses is governed by a Hamiltonian like that in Eq.~(\ref{eq:ham}) and this dynamics is in turn probed by emitted particles, either by employing an additional probe pulse, or  by the same pulse. The emitted particles do not have to be photoelectrons. They can be, e.g., Auger electrons or photons. They all serve as a probe of the few-level system coupled by the pump pulse.

In the case of resonant Auger decay of an atom in a free electron laser field studied in Refs.~\cite{Rohringer08,Liu,Demekhin11SFatom}, a coherent high-energy laser pulse of resonant carrier frequency couples the ground state and a core-excited electronic state. The latter state decays by emitting an Auger electron. The quantum motion of this two-level system is described by a $2\times2$ Hamiltonian similar to that considered here (see  Eq.~(34) of Ref.~\cite{Demekhin11SFatom}). The main difference to the Hamiltonian~(\ref{eq:ham}) is the presence of a time-independent Auger rate $\Gamma_{A}$ of the core-excited state in addition to the leakage by ionization to the continuum present in Eq.~(\ref{eq:ham}). We thus have to add to  $ -\frac{ i }{2}\Gamma g^2(t) $  describing the leakage in Eq.~(\ref{eq:ham}) the term $-\frac{i}{2}\Gamma_{A}$ describing the Auger decay. The time-independent rate $\Gamma_{A}$ can be substantial \cite{Rohringer08,Liu,Demekhin11SFatom} and is then not negligible compared to $\Delta\,g(t)$ at least at the very beginning and very end of an intense pulse. However, whenever during the pulse the field-induced coupling $\Delta\,g(t)$ between the two states becomes larger than the Auger decay width  $\Gamma_{A}$, the above discussed scenario of decoupled resonances can be applied and dynamic interference takes place. Indeed, multiple-peak patterns in the Auger  spectrum are found in Refs.~\cite{Rohringer08,Demekhin11SFatom}, but hitherto not interpreted. In view of the present results,  these patterns can be understood in terms of dynamic interference. In those cases where photons are emitted from the decoupled resonances, e.g., X rays,  it is clear that the respective emission spectra of atoms exposed to coherent intense pulses will also exhibit dynamic interference effects. The only difference  will be that a term $-\frac{i}{2}\Gamma_{X}$ will have to be added to Hamiltonian (\ref{eq:ham}) to account for the relaxation of the intermediate state via spontaneous emission.



\begin{thebibliography}{38}
\expandafter\ifx\csname natexlab\endcsname\relax\def\natexlab#1{#1}\fi
\expandafter\ifx\csname bibnamefont\endcsname\relax
  \def\bibnamefont#1{#1}\fi
\expandafter\ifx\csname bibfnamefont\endcsname\relax
  \def\bibfnamefont#1{#1}\fi
\expandafter\ifx\csname citenamefont\endcsname\relax
  \def\citenamefont#1{#1}\fi
\expandafter\ifx\csname url\endcsname\relax
  \def\url#1{\texttt{#1}}\fi
\expandafter\ifx\csname urlprefix\endcsname\relax\def\urlprefix{URL }\fi
\providecommand{\bibinfo}[2]{#2}
\providecommand{\eprint}[2][]{\url{#2}}

\bibitem{B1}
M.V. Fedorov,  \emph{Atomic and free electrons in a strong light field} (World Scientific, Singapore, 1997).

\bibitem{B2}
R. Loudon,  \emph{The quantum theory of light} (Oxford U. P., Oxford, 2000), 3rd ed.

\bibitem{B3}
N.B. Delone and V.P. Krainov, \emph{Multiphoton Processes in Atoms} (Springer, Heidelberg, 2000), 2nd ed.

\bibitem{B4}
C. Gerry and P. Knight,  \emph{Introductory quantum optics} (Cambridge U. P., Cambridge, 2004).

\bibitem{NIMROD}
N. Moiseyev,  \emph{Non-Hermitian Quantum Mechanics} (Cambridge U. P., Cambridge, 2011).

\bibitem{ZEWAIL}
A.H. Zewail, \emph{Femtochemistry, vol. I and II} (World Scientific, Singapore, 1994).

\bibitem{KRAUSZ}
F.\,Krausz\,and\,M.\,Ivanov, Rev.\,Mod.\,Phys. \textbf{81}, 163 (2009).

\bibitem{THbp1}
S. Gu\'{e}rin and H. R. Jauslin, Adv. Chem. Phys. \textbf{125}, 147 (2003).

\bibitem{THbp2}
E. Gamaly, \emph{Femtosecond Laser-Matter Interaction: Theory, Experiments and Applications} (Pan Stanford Publishing Pte. Ltd., Singapore, 2011).

\bibitem{THbp3}
B.W. Shore, \emph{Manipulating Quantum Structures Using Laser Pulses} (Cambridge U. P., New York, 2011).

\bibitem{AT}
S.H.\,Autler and C.H.\,Townes,\,Phys.\,Rev.\,\textbf{100},\,703\,(1955).

\bibitem{AUI1}
K. Rz\c{a}\.{z}ewski, Phys. Rev. A  \textbf{28}, 2565 (1983).

\bibitem{MPI1}
D. Rogus and M. Lewenstein, J. Phys. B \textbf{19}, 3051  (1986).

\bibitem{MPI2}
C. Meier and V. Engel,  Phys. Rev. Lett.  \textbf{73}, 3207 (1994).

\bibitem{Rongqing}
C. Rongqing, X. Zhizhan, S. Lan, Y. Guanhua, and Z. Wenqui, Physical Review A  \textbf{44}, 558 (1991).

\bibitem{Sussman11}
B.J. Sussman,  { Am. J. Phys.} \textbf{79}, 477 (2011).

\bibitem{highharm1}
G. Sansone, \emph{et al.},  {Science} \textbf{314}, 443 (2006).

\bibitem{highharm2}
E. Goulielmakis,  \emph{et. al.},  {Science} \textbf{320}, 1614 (2008).

\bibitem{FLASH}
W. Ackermann, \emph{et al.},  {Nature photonics} \textbf{1}, 336  (2007).

\bibitem{FERMI}
Home page of  FERMI at Elettra in Trieste, Italy, http://www.elettra.trieste.it/FERMI/.

\bibitem{DynIntLETT}
Ph.V. Demekhin   and L.S. Cederbaum, Phys. Rev. Lett. \textbf{108}, 253001 (2012).

\bibitem{NOTOURS}
K. Toyota, O.I. Tolstikhin, T. Morishita,  S. Watanabe, Phys.  Rev. A \textbf{76}, 043418 (2007); \emph{ibid.} \textbf{78}, 033432 (2008).


\bibitem{EXPT}
R.R. Jones, D.W. Schumacher, and P.H. Bucksbaum, Phys. Rev. A \textbf{47}, R49 (1993); J.G. Story, D.I. Duncan, and T.F. Gallager, Phys. Rev. Lett. \textbf{70}, 3012 (1993); R.B. Vrijen, J.H. Hoogenraad, H.G. Muller, and L.D. Noordam, \emph{ibid.} \textbf{70}, 3016 (1993).

\bibitem{jones}
R.R. Jones, Phys. Rev. Lett. \textbf{74}, 1091 (1995); \emph{ibid.} \textbf{75}, 1491 (1995).

\bibitem{Chiang10}
Y.-C. Chiang,  Ph.V. Demekhin, A.I. Kuleff, S. Scheit, and L.S. Cederbaum,  {Phys. Rev. A} \textbf{81}, 032511 (2010).


\bibitem{Demekhin11SFatom}
Ph.V. Demekhin {and} L.S. Cederbaum, Phys. Rev. A \textbf{83}, 023422 (2011).

\bibitem{MolRaSfPRL}
L.S. Cederbaum,  Y.-C. Chiang, Ph.V. Demekhin, and N. Moiseyev,  {Phys. Rev. Lett.} \textbf{106}, 123001 (2011).

\bibitem{DemekhinCO}
Ph.V. Demekhin, Y.-C.  Chiang, and L.S. Cederbaum, {Phys. Rev. A.} \textbf{84}, 033417 (2011).

\bibitem{DemekhinICD}
Ph.V. Demekhin, S.D. Stoychev, A.I. Kuleff, and L.S. Cederbaum, Phys. Rev. Lett \textbf{107}, 273002  (2011).

\bibitem{Cederbaum81}
L.S. Cederbaum  and W. Domcke,  {J. Phys. B.} \textbf{14}, 4665 (1981).

\bibitem{Domcke91}
W. Domcke, {Phys. Rep.} \textbf{208},   97  (1991).

\bibitem{Sun}
Y.-P. Sun, J.-C. Liu, C.-K. Wang, and F. K. Gel'mukhanov, Phys. Rev. A  \textbf{81}, 013812 (2010).

\bibitem{statphase}
N. Bleistein and R. Handelsman,   \emph{Asymptotic Expansions of Integrals.} (Dover, New York, 1975).


\bibitem{Rohringer08}
N. Rohringer and R. Santra, Phys. Rev. A \textbf{77}, 053404 (2008).

\bibitem{Liu}
J.-C. Liu, Y.-P. Sun, C.-K. Wang, H. {\AA}gren, and F. K. Gel'mukhanov, Phys. Rev. A \textbf{81}, 043412 (2010).


\end{thebibliography}
\end{document}